\documentclass[amsfonts,amsmath,prd,preprint,nofootinbib]{revtex4}

\usepackage{graphicx}
\usepackage[T1]{fontenc}
\usepackage{lmodern}
\usepackage[breaklinks=true]{hyperref}

\usepackage{amsfonts,amsmath}
\usepackage{epsfig,bbm,cancel,ulem}
\usepackage{xcolor}
\usepackage{wasysym}
\usepackage[mathscr]{eucal}
\usepackage{multirow}
\usepackage{enumitem}
\usepackage{appendix}
\usepackage{slashed}

\newcommand{\beq}{\begin{equation}}
\newcommand{\eeq}{\end{equation}}

\newcommand{\bsp}{\begin{split}}

\begin{document}

\title{Bogomol'nyi Equations for Kruglov Strings}

\author{I.~Prasetyo}
\email{ilham.prasetyo@sampoernauniversity.ac.id}
\thanks{ORCID: \href{https://orcid.org/0000-0002-0879-7777}
{0000-0002-0879-7777}}
\affiliation{Department of Mechanical Engineering, FET,
Sampoerna University, Jakarta 12780, Indonesia}
\affiliation{General Education Unit, FASE,
Sampoerna University, Jakarta 12780, Indonesia}

\author{U.~Ubaydillah}
\email{ury.ubaydillah@ui.ac.id}
\thanks{ORCID: \href{https://orcid.org/0009-0007-1251-9669}
{0009-0007-1251-9669}}
\affiliation{Departemen Fisika, FMIPA
Universitas Indonesia, Depok 16424, Indonesia}

\author{H.~S.~Ramadhan}
\email{hramad@sci.ui.ac.id}
\thanks{{\bf Corresponding author.}}
\thanks{ORCID: \href{https://orcid.org/0000-0003-0727-738X}
{0000-0003-0727-738X}}
\affiliation{Departemen Fisika, FMIPA
Universitas Indonesia, Depok 16424, Indonesia}

\begin{abstract}
We construct the Bogomol'nyi equations for Abelian gauge--Higgs vortices in which the Maxwell gauge sector is replaced by Kruglov nonlinear electrodynamics, a power-law family that interpolates between Maxwell theory, Born--Infeld electrodynamics, and exponential electrodynamics, characterized by a dimensionless exponent $\sigma$. Using the stressless method, we derive a pair of first-order equations directly from the vanishing of the spatial stress tensor, without assuming the Higgs potential \textit{a priori}. For generic $\sigma$, the gauge and Higgs sectors are coupled through an implicit algebraic relation. We therefore introduce a constitutive map $\Phi\left(Y;\sigma\right)$ and analyze its monotonicity and range to determine the conditions for a smooth admissible Bogomol'nyi branch. For $\sigma>1/2$, the constitutive map is strictly monotonic and unbounded, whereas for $0<\sigma<1/2$ it possesses a finite maximum; the marginal case $\sigma=1/2$ is bounded. These properties yield explicit bounds on the nonlinear parameter $\beta$ for the latter cases. We further obtain closed-form constitutive relations, BPS potentials, and gauge-field equations for six representative values of $\sigma$, spanning linear, quadratic, and cubic algebraic structures. The corresponding vortex profiles are then computed numerically. The resulting BPS string tension is purely topological, $\mu_{\rm BPS}=2\pi n$, independent of both $\sigma$ and $\beta$.
\end{abstract}

\maketitle


%
\section{Introduction}
\label{sec1}

Nonlinear electrodynamics (NLED) has attracted significant attention in modern theoretical physics, particularly due to its natural appearance in the low-energy effective regimes of String Theory~\cite{Fradkin:1985qd,Leigh:1989jq,Seiberg:1999vs}. Originally proposed by Born and Infeld to regularize the divergent self-energy of the electron~\cite{Born:1934gh}, NLED has found applications in strong-field astrophysics, such as in the environments surrounding magnetars and charged black holes~\cite{Kruglov2015}. NLED provides an effective description of electromagnetic fields in the strong-field regime, while recovering Maxwell electrodynamics in the weak-field limit. When coupled to gravity, the nonlinear electromagnetic stress-energy tensor can generate effective negative pressure and provide a possible mechanism for inflationary expansion without modifying General Relativity~\cite{Garcia-Salcedo:2002bys,Novello:2003kh,Vollick:2008dx}. 
More recently, NLED has also been investigated as a
theoretical framework for describing vacuum birefringence~\cite{Kruglov2015,Kruglov2017}.

The Born--Infeld (BI) theory modifies Maxwell electrodynamics by replacing the standard quadratic Lagrangian with a square-root form that introduces a characteristic field-strength scale~\cite{Born:1934gh}. A natural extension of the BI framework is provided by the Kruglov model~\cite{Kruglov2017}. In this model, the square-root structure of the Born--Infeld Lagrangian is generalized to a power-law form controlled by a dimensionless parameter $\sigma$, together with the nonlinear scales $\beta$ and $\gamma$. The model contains, in its spectrum, Maxwell electrodynamics as the $\sigma=1$ limit, and reduces to the Born--Infeld theory for $\sigma=1/2$ (with $\beta=\gamma$). In the limit
$\sigma\to\infty$, it approaches exponential electrodynamics. For $\sigma<1$, the electric field remains finite at the origin, with
$E(0)=\sqrt{2\sigma/\beta}$. For $\beta\neq\gamma$, the model admits vacuum birefringence.~\cite{Kruglov:2007bh}.

In the Abelian--Higgs model, the coupling of a $U(1)$ gauge field to a symmetry-breaking scalar field gives rise to topological vortex solutions, which appear in various contexts including superconductivity and cosmic strings~\cite{Abrikosov:1956sx,Nielsen:1973cs}. When the gauge sector governing the Abelian--Higgs vortices is extended from Maxwell to NLED, the nonlinear gauge-field self-interactions modify the vortex profiles and energy-density
distribution. Born--Infeld-Higgs models provide a natural example of such a nonlinear extension, and vortex solutions in these models have been studied in both planar and generalized settings~\cite{Shiraishi:1990zi,Bazeia:2011bz}. More general Born--Infeld-Higgs theories have subsequently been considered by introducing additional Higgs-dependent functions into the gauge and scalar sectors, leading to a broader class of vortex solutions~\cite{Casana:2015bea}. 

At a critical relation between the gauge and scalar couplings, the Abelian--Higgs vortices admit a Bogomol'nyi--Prasad--Sommerfield (BPS)
limit, in which the second-order Euler--Lagrange equations reduce to a set of coupled first-order equations~\cite{Bogomolny:1975de,Prasad:1975kr}.
The resulting solutions saturate the minimum energy within a fixed topological sector, with the string tension scaling linearly with the winding number. While the Bogomol'nyi equations are well-established for the canonical Maxwell--Higgs theory, their construction becomes considerably nontrivial when the gauge and scalar sectors are described by nonlinear or non-canonical kinetic terms. In such theories, the appropriate first-order equations are not generally evident from the standard Bogomol'nyi decomposition, motivating the development of systematic methods for constructing BPS equations beyond the canonical setting. Over the last two decades, several mechanisms have been developed in this way, including the first-order formalisms~\cite{Bazeia:2017nas}, the on-shell method~\cite{Atmaja:2014fha}, and the BPS Lagrangian method~\cite{Atmaja:2015umo}, with applications to generalized vortices and other  topological defects~\cite{Atmaja:2015umo,Atmaja:2018cod,Gunawan:2024pvc}.

A particularly useful approach to constructing Bogomol'nyi equations is based on the {\it stressless condition}, which requires the spatial components of the energy--momentum tensor to vanish. For static and cylindrically symmetric configurations, conservation of the energy--momentum tensor relates the radial and angular stresses, allowing the BPS sector to be
identified through the conditions $T_{rr}=T_{\theta\theta}=0$. This approach was originally employed in the analysis of Abelian--Higgs vortices by de Vega and Schaposnik~\cite{deVega:1976xbp} and was
later developed in the context of generalized vortex theories~\cite{Bazeia:2017nas}. The stressless condition has also been
applied to more general topological defects, including non-Abelian Alice strings~\cite{Acalapati:2023rxf} and Dirac--Born--Infeld cosmic strings~\cite{Ramadhan:2025rxj}, where it provides an independent
consistency check of the corresponding first-order equations. In this framework, the first-order equations are obtained directly from the stressless conditions and can subsequently be checked for consistency with the second-order equations of motion.

Motivated by these developments, in this work we investigate BPS states in an Abelian gauge--Higgs theory whose Maxwell gauge sector is replaced by Kruglov nonlinear electrodynamics. We employ the stressless condition as a direct method for constructing the corresponding Bogomol'nyi equations, without relying on the standard Bogomol'nyi decomposition. This work is organized as follows. In Sec.~\ref{sec2}, we establish the field equations for static, cylindrically symmetric string configurations in the Abelian Kruglov-Higgs action. In Sec.~\ref{sec3} we impose the vanishing of the spatial stresses and derive the resulting first-order equations and the corresponding BPS tension. Owing to the nonlinear structure of the gauge sector, Sec.~\ref{sec4} is devoted to a rigorous mathematical analysis of the constraints on the parameter $\sigma$. In Sec.~\ref{sec5}, we derive the Bogomol'nyi equations obtain explicit BPS solutions for several representative admissible values of $\sigma$. Finally, we summarize our results in Sec.~\ref{sec6}.


\section{Kruglov--Higgs model}
\label{sec2}

We consider an Abelian gauge--Higgs theory in which the Maxwell gauge sector is replaced by Kruglov nonlinear electrodynamics. The action is
given by~\cite{Kruglov2017}\footnote{Recently, one of us discussed the photon propagation black hole imaging in this type of NLED~\cite{Ramadhan:2026ekb}.}
\begin{subequations}
\begin{align}
    \mathcal{S}
    &=\int d^4x \sqrt{-g}
    \Big[\mathcal{L}_{\text{Higgs}}
    +\mathcal{L}_{\text{Kruglov}}\Big],
    \\
    \mathcal{L}_{\text{Higgs}}
    &=|D_\mu \phi|^2 - V\left(|\phi|\right),
    \\
    \mathcal{L}_{\text{Kruglov}}
    &=
    {1\over \beta}\Bigg[
    1-\Bigg(
    1+\frac{\beta \mathcal{F}}{\sigma}
    -\frac{\beta\gamma \mathcal{G}^2}{2\sigma}
    \Bigg)^\sigma
    \Bigg],
    \label{eq:LKruglov}
\end{align}
\end{subequations}
where we adopt the metric signature $(+,-,-,-)$ and define
\[
D_\mu \phi=\partial_\mu\phi-ieA_\mu\phi,
\qquad
F_{\mu\nu}=\partial_\mu A_\nu-\partial_\nu A_\mu,
\]
together with
\[
\tilde{F}^{\mu\nu}
=
\frac{\varepsilon^{\mu\nu\kappa\rho}}
{2\sqrt{-g}}F_{\kappa\rho},
\qquad
\mathcal{F}
=
\frac{1}{4}F_{\mu\nu}F^{\mu\nu},
\qquad
\mathcal{G}
=
\frac{1}{4}F_{\mu\nu}\tilde{F}^{\mu\nu}.
\]
Here, $\beta$ and $\gamma$ have dimensions of
$[\mathrm{length}]^4$, while $\sigma$ is dimensionless~\cite{Kruglov2017}. The model contains Maxwell electrodynamics in the $\sigma=1$ limit and reduces to the Born--Infeld theory for $\sigma=1/2$ when $\beta=\gamma$. In the limit $\sigma\to\infty$, it approaches exponential electrodynamics. For $\sigma<1$, the electric field remains finite at the origin, with
$E(0)=\sqrt{2\sigma/\beta}$~\cite{Kruglov2017}.
Throughout this work, we set $\gamma=\beta/2\sigma$ for the reason that we shall reveal later in Sec.~\ref{sec6}.

We consider a static, cylindrically symmetric string configuration described by the standard Nielsen-Olesen-type ansatz~\cite{Nielsen:1973cs}:
\begin{subequations}
\begin{align}
    A_t&=A_r=0,
    \qquad
    A_\theta=\frac{n-a(r)}{e},
    \\
    \phi&=f(r)\exp\left(i n \theta\right),
    \\
    g_{\mu\nu}&=\mathrm{diag}\left(1,-1,-r^2,-1\right),
\end{align}
\end{subequations}
where $n\in\mathbb{Z}$ denotes the winding number. The only independent nonvanishing component of the electromagnetic field strength is
\[
F_{r\theta}=-\frac{a'}{e},
\]
which gives the magnetic field along the string,
\[
B=\frac{F_{r\theta}}{r}=-\frac{a'}{er}.
\]
The electric field vanishes for this configuration, and therefore
\begin{equation}
\mathcal{F}=\frac{B^2}{2}=\frac{1}{2}\left(\frac{a'}{er}\right)^2,
\qquad\mathcal{G}=0.
\end{equation}

Consequently, the action reduces to
\begin{eqnarray}
    S&=&\int r\,\mathcal{L}_{\mathrm{eff}}\,dt\,dr\,d\theta\,dz,\nonumber\\
    \mathcal{L}_{\mathrm{eff}}&=&-f'^2
    -\frac{a^2f^2}{r^2}-V(f)\nonumber\\
    &&\quad+{1\over\beta}\Bigg[1-\Bigg(1+\frac{\beta}{\sigma}
    \frac{a'^2}{2e^2r^2}\Bigg)^\sigma\Bigg].\label{eq:Seff}
\end{eqnarray}
The vortex profiles obey the boundary conditions
\begin{align}
    f(0)&=0,& f(\infty)&=1,\nonumber\\
    a(0)&=n,& a(\infty)&=0.\label{boundary}
\end{align}

\section{First-order formalism}
\label{sec3}

\subsection{Stressless method}

The Bogomol'nyi equations can be constructed by imposing the stressless condition on the energy--momentum tensor. This approach was first employed in the analysis of Abelian--Higgs vortices by de Vega and Schaposnik~\cite{deVega:1976xbp} and has subsequently been applied to generalized vortex theories~\cite{Bazeia:2017nas}. For the static and cylindrically symmetric configuration considered here, the radial component of the conservation law,
\[
\nabla_\mu T^\mu{}_\nu=0,
\]
takes the form
\begin{equation}
\frac{d}{dr}\left(rT^r{}_r\right)=T^\theta{}_\theta.
\end{equation}
A sufficient condition for satisfying this relation is the stressless condition
\begin{equation}
\label{stressless}
T^r{}_r=T^\theta{}_\theta=0.
\end{equation}
These two conditions provide the first-order constraints from which
the Bogomol'nyi equations can be constructed.

The energy--momentum tensor can be written as~\cite{Cordero:2007rw}
\begin{align}
T_{\mu\nu}&=T_{\mu\nu}^{\text{Higgs}}+
T_{\mu\nu}^{\text{Kruglov}},\nonumber\\
T_{\mu\nu}^{\text{Higgs}}
&=
g_{\mu\nu}\mathcal{L}_{\text{Higgs}}
-\left(D_{\mu}\phi\right)^*\left(D_{\nu}\phi\right)
-\left(D_{\nu}\phi\right)^*\left(D_{\mu}\phi\right),
\nonumber\\
T_{\mu\nu}^{\text{Kruglov}}&=g_{\mu\nu}\mathcal{L}_{\text{Kruglov}}+\left(1+\frac{\beta\mathcal{F}}{\sigma}\right)^{\sigma-1}\left(F_{\mu\alpha}F_{\nu\beta}g^{\alpha\beta}-\gamma\mathcal{G}^2g_{\mu\nu}\right).
\end{align}

After substituting the ansatz and raising one index of the energy--momentum tensor, its nonvanishing diagonal components can be expressed in terms of the effective Lagrangian as
\begin{subequations}
\begin{align}
T^t_{\ t}
&=
\mathcal{L}_{\text{eff}}
=
T^z_{\ z},
\\
T^r_{\ r}
&=
\mathcal{L}_{\text{eff}}
+
K
+
2f'^2,
\\
T^\theta_{\ \theta}
&=
\mathcal{L}_{\text{eff}}
+
K
+
\frac{2a^2f^2}{r^2},
\end{align}
\end{subequations}
where, for convenience, we define
\begin{align}
\mathcal{L}_{\text{eff}}&\equiv-f'^2-\frac{a^2f^2}{r^2}-V+\frac{1}{\beta}\left(1-Y^\sigma\right),\nonumber\\
K&\equiv\frac{a'^2}{\left(er\right)^2}Y^{\sigma-1},\nonumber\\
Y&\equiv1+\frac{\beta}{\sigma}\frac{a'^2}{2\left(er\right)^2}.
\label{eq:X}
\end{align}

The two stressless conditions in Eq.~\eqref{stressless} are equivalently expressed through their sum and difference,
\begin{subequations}
\begin{align}
T^r_{\ r}+T^\theta_{\ \theta}&=0,\label{T+T}\\
T^r_{\ r}-T^\theta_{\ \theta}&=0.\label{T-T}
\end{align}
\end{subequations}
These two combinations provide the first-order constraints that initiate the construction of the Bogomol'nyi equations. 

Eq.~\eqref{T-T} eliminates both the potential and the nonlinear gauge contribution, giving
\begin{equation}
f'=\pm\frac{af}{r}.
\label{eq:f'}
\end{equation}
This relation already takes the form of one of the first-order Bogomol'nyi equations. The remaining first-order relation for the gauge profile is obtained by requiring consistency with the Euler--Lagrange equations.

Combining Eq.~\eqref{T+T} with~\eqref{eq:f'} gives
\begin{eqnarray}
\frac{a'^2}{\left(er\right)^2}Y^{\sigma-1}-V+\frac{1}{\beta}\left(1-Y^\sigma\right)&=&0.
\end{eqnarray}
Using the definition of $Y$ in Eq.~\eqref{eq:X}, the gauge-field derivative can be eliminated in favor of $Y$. This yields
\begin{equation}
V=\frac{\left(2\sigma-1\right)Y^\sigma-2\sigma Y^{\sigma-1}+1}{\beta}.
\label{eq:V}
\end{equation}
Once the relation between $Y$ and the Higgs profile $f$ is established, the potential can be specified.

To determine the remaining first-order relation, we require consistency with the Euler--Lagrange equations. Variation of the effective action~\eqref{eq:Seff} with respect to $f$ gives
\begin{equation}
\left(2rf'\right)'-\frac{2a^2f}{r}-rV_f=0,
\end{equation}
where $V_f\equiv\partial V/\partial f$. Substituting Eq.~\eqref{eq:f'} into this equation, and taking the positive root, gives
\begin{equation}
V_f=
\frac{2 fa'}{r}.
\label{eq:V'}
\end{equation}
Since the left-hand side depends only on $f$, this relation shows that $a'/r$ must likewise be expressible as a function of $f$.

The Euler--Lagrange equation for the gauge profile follows from variational action with Lagrangian density Eq.~\eqref{eq:Seff} with respect to $a$,
\begin{equation}
\left(
\frac{a'}{e^2r}Y^{\sigma-1}
\right)'
=
\frac{2af^2}{r}.
\end{equation}
Using Eq.~\eqref{eq:f'}, it can be integrated once to give
\begin{equation}
\frac{a'}{e^2r}Y^{\sigma-1}
- f^2
=
C_a,
\end{equation}
with $C_a$ an integration constant. The boundary conditions~\eqref{boundary}, along with the requirement that the magnetic field vanish asymptotically, imply $C_a=-1$. Consequently,
\begin{equation}
\frac{a'}{r}
=
 e^2\left(f^2-1\right)Y^{1-\sigma}.
\label{eq:a'}
\end{equation}

Combining Eq.~\eqref{eq:a'} with the definition of $Y$ in Eq.~\eqref{eq:X} gives an algebraic relation between $Y$ and the Higgs profile,
\begin{equation}
\frac{\beta e^2}{2\sigma}\left(f^2-1\right)^2
=\left(Y-1\right)Y^{2\sigma-2}.\label{eq:findX}
\end{equation}
Eq.~\eqref{eq:findX} is central to the subsequent analysis in Sec.~\ref{sec4}. For a given value of $\sigma$, it determines the admissible branches of $Y$ as a function of $f$, from which $a'/r$ and $V(f)$ can be obtained through Eqs.~\eqref{eq:a'} and~\eqref{eq:V}, respectively. In contrast, the Higgs first-order equation in Eq.~\eqref{eq:f'} is independent of $Y$.

The algebraic structure of Eq.~\eqref{eq:findX} also imposes nontrivial constraints on the allowed values of the model parameters. In particular, since $\sigma\neq0$ and $\beta\neq0$, the sign and range of $Y$ depend on the sign of $\beta$. For $\beta>0$, Eq.~\eqref{eq:X} implies $Y\geq1$, whereas for $\beta<0$ the admissible range of $Y$ requires a separate analysis. These restrictions, together with the requirement that the resulting potential and profile functions remain physically and mathematically well defined, will be examined in Sec.~\ref{sec4}.

\subsection{BPS tension}

The string tension, defined as the static energy per unit length of the string, can be evaluated in the first-order sector. Since $T^r_{\ r}=0$, the effective Lagrangian evaluated on the first-order
solutions satisfies
\begin{align}
\mathcal{L}_{\text{eff}}\big|_{\text{BPS}}
&=
-K-2f'^2
\nonumber\\
&=
-\frac{a'^2}{\left(er\right)^2}Y^{\sigma-1}
-2f'^2.
\end{align}
Using Eq.~\eqref{eq:f'}, the second term becomes
\begin{align}
-2f'^2
&=
-2\frac{af f'}{r},
\end{align}
while first term gives, using Eq.~\eqref{eq:a'},
\begin{align}
-\frac{a'^2}{\left(er\right)^2}Y^{\sigma-1}
&=
-\frac{ a'}{r}\left(f^2-1\right).
\end{align}
Consequently,
\begin{align}
\mathcal{L}_{\text{eff}}\big|_{\text{BPS}}
&=-\frac{1}{r}\frac{d}{dr}\left[a\left(f^2-1\right)\right].
\end{align}

Since the configuration is static and independent of $\theta$ and
$z$, the string tension is
\begin{eqnarray}
\mu_{\text{BPS}}
&=&
\int r T^t_{\ t}\big|_{\text{BPS}}\,dr\,d\theta
\nonumber\\
&=&
-2\pi
\int_0^\infty
r\mathcal{L}_{\text{eff}}\big|_{\text{BPS}}\,dr
\nonumber\\
&=&
2\pi 
\left[
a\left(f^2-1\right)
\right]_0^\infty
\nonumber\\
&=&
2\pi  n,
\end{eqnarray}
where the boundary conditions in Eq.~\eqref{boundary} have been used. The BPS tension is therefore determined solely by the topological winding number and is independent of the nonlinear parameters $\sigma$ and $\beta$, consistent with the general topological saturation expected of Bogomol'nyi-Prasad-Sommerfield solitons~\cite{Bogomolny:1975de,Prasad:1975kr}.

\section{Structure of the constitutive map and existence of an admissible Bogomol'nyi branch}
\label{sec4}


The second Bogomol'nyi equation, Eq.~\eqref{eq:a'}, determines the gauge-field profile through the auxiliary variable $Y$, which is implicitly coupled to the Higgs profile via Eq.~\eqref{eq:findX}. In general, this algebraic relation cannot be inverted in closed form for arbitrary values of $\sigma$. By the Abel--Ruffini theorem~\cite{Stewart2015}, polynomial equations of degree five or higher are generally unsolvable by radicals; hence, for generic choices of $\sigma$, expressing $Y(f)$ explicitly in terms of elementary functions is algebraically impossible. While obtaining $Y(f)$ analytically is intractable compared to evaluating the inverse mapping $f(Y)$, the Implicit Function Theorem ensures that working with $a'/r = F\left[Y(f)\right]$ is functionally equivalent to parameterizing the system via $a'/r = F\left[f(Y)\right]$. Therefore, before proceeding to numerical solutions, it is instructive to establish the existence and uniqueness of an admissible constitutive branch $Y=Y(f)$ directly from the intrinsic properties of the map defined by Eq.~\eqref{eq:findX}.

\subsection{Domain of the auxiliary variable}

First, the Lagrangian~\eqref{eq:LKruglov} remains real for arbitrary non-integer $\sigma$ only if
\begin{equation}
\label{eq:Y>0}
Y>0.
\end{equation}
Assuming $\beta>0$, for $\sigma<0$ then Eq.~\eqref{eq:Y>0} places a constraint on $B$:
\[
1-{\beta\over2|\sigma|}B^2>0;
\]
magnetic field must be bounded from above:
\begin{equation}
B^2<{|\sigma|\over2\beta}.
\end{equation}
Boundedness of
$\mathcal L_{\text{Kruglov}}$ from below further restricts 
\begin{equation}
    0<Y<1.
\end{equation}
For $\sigma>0$, by contrast, the definition of $Y$, Eq.~\eqref{eq:X}, forces 
\begin{equation}
\label{boundY}
    Y>1.
\end{equation}

\subsection{The constitutive map $\Phi(Y;\sigma)$}

Within the physical domain, Eq.~\eqref{eq:findX} can be written 
\begin{equation}
\label{eq:constitutive}
\Phi\left(Y;\sigma\right)=G(f),
\end{equation}
where
\begin{equation}
\label{eq:Phi}
\Phi\left(Y;\sigma\right)\equiv\left(Y-1\right)Y^{2\sigma-2}
\end{equation}
is called {\it the constitutive map}, and
\begin{equation}
G(f)\equiv\frac{\beta e^2}{2\sigma}\left(f^2-1\right)^2.
\end{equation}
Eq.~\eqref{eq:constitutive} shows that the nonlinear coupling between the magnetic field and the Higgs field is encoded in the constitutive map $\Phi(Y;\sigma)$. Consequently, the existence and uniqueness of an admissible Bogomol'nyi branch are determined entirely by the properties of this function. 

In other words, if both $\Phi\left(Y;\sigma\right)$ and $G(f)$ are monotonically increasing or decreasing, the relation between $Y$ and $f$ is one-to-one. Since we know from the boundary conditions that $G$ is monotonically decreasing (increasing only if $\beta\sigma<0$) as $f$ increases from $f=0$ to $f=1$, we examine $\Phi(Y;\sigma)$. If $\Phi\left(Y;\sigma\right)$ is not monotonically increasing or decreasing, then there exists critical values of $Y=Y_c$ such that $\partial\Phi\left(Y;\sigma\right)/\partial Y\equiv \Phi'=0$.

The critical value $Y_c$ happens for
\[
\Phi'=0\rightarrow\quad Y_c\in\Big\{0,{2\sigma-2\over2\sigma-1}\Big\}, \quad\sigma\neq1/2.
\]
Disregarding the trivial $Y_c=0$, we have
\[
Y_c-1={1\over1-2\sigma}.
\]
The location of the stationary point depends entirely on whether $\sigma<1/2$ or $\sigma>1/2$. The case for $\sigma=1/2$ will be treated independently. 

For $\sigma>1/2$ one finds $Y_c<1$, so that the stationary point lies outside the physical domain~\eqref{boundY}. Consequently, 
\[
\Phi'\left(Y;\sigma>1/2\right)>0,
\]
and the constitutive map is strictly monotonic. Since  
\[
\Phi(1;\sigma)=0,\quad\text{and}\quad\Phi(Y\rightarrow\infty;\sigma>1/2)\sim Y^{2\sigma-1},
\]
the function increases continuously. The constitutive map establishes a one-to-one correspondence between $Y\in\left(1,\infty\right)$ and positive values of $G(f)$. The map is thus uniquely invertible over the entire physical domain.

For $0<\sigma<1/2$ the situation changes qualitatively. Here,
\[
Y_c>1,
\]
so that the stationary point lies within the physical domain. The constitutive map starts from $Y=1$, then increases until it reaches a maximum at
$Y=Y_c$, and subsequently decreases towards zero as
$Y\rightarrow\infty$.
The maximum value,
\begin{equation}
\label{eq:Phimax}
\Phi_{\max}=
\Phi(Y_c;\sigma)=\frac1{|2\sigma-1|}\bigg(\frac{2\sigma-2}{2\sigma-1}\bigg)^{2\sigma-2},
\end{equation}
marks the largest value attainable by the map.
Consequently,
\[
0<G(f)<\Phi_{\max},
\]
which generally admits two distinct algebraic solutions for $Y$, corresponding to two possible constitutive branches.

The marginal case $\sigma=1/2$ interpolates between these two regimes. Here 
\[
\Phi\left(Y;1/2\right)=1-{1\over Y},
\]
which remains monotonic throughout the physical domain but approaches the finite limit
\[
\sup_{Y>1}\Phi\left(Y\rightarrow\infty;1/2\right)=1.
\]

\subsection{Existence of an admissible constitutive branch}

The boundary conditions~\eqref{boundary} imply
\[
0\le G(f)\le G_{\text{max}}\equiv{\beta e^2\over2\sigma}.
\]
A constitutive branch connecting the vortex core to spatial infinity exists provided the source term never exceeds the largest value attainable by the constitutive map,
\begin{equation}
\label{eq:existence}
\frac{\beta e^2}{2\sigma}
\le
\sup_{Y>1}\Phi\left(Y;\sigma\right).
\end{equation}

Whenever Eq.~\eqref{eq:existence} is satisfied, the physically relevant branch is the one continuously connected to the vacuum value $Y=1$. Along this branch the stationary point $Y_c$ is never reached, so that $\Phi'(Y)\neq0$ everywhere. The Implicit Function Theorem then guarantees that the constitutive relation $Y=Y(f)$ is smooth throughout the interval $0\le f\le1$.

If instead
\begin{equation}
\frac{\beta e^2}{2\sigma}>\sup_{Y>1}\Phi(Y;\sigma),
\end{equation}
the constitutive branch terminates before reaching the vortex core, indicating that Eq.~\eqref{eq:findX} admits no real solution over the entire physical interval.

Eq.~\eqref{eq:existence} thus immediately separates the parameter space into
two different sectors. For $\sigma>\frac12$, the constitutive map is unbounded,
\[
\sup_{Y>1}\Phi\left(Y;\sigma>1/2\right)=\infty,
\]
implying Eq.~\eqref{eq:existence} is automatically satisfied for any
positive values of $\beta$ and $e$. For $0<\sigma<\frac12$, the map possesses the finite upper bound
$\Phi_{\max}$, leading to the nontrivial constraint
\begin{equation}
\beta<\frac{2\sigma}{e^2}
\Phi_{\max}.
\label{eq:sigmalt1p2}
\end{equation}
Lastly, for the marginal case $\sigma=1/2$, the condition reduces to
\begin{equation}
\label{38}
\beta\leq\frac{1}{e^2}.
\end{equation}

We emphasize that the above analysis concerns only the algebraic constitutive relation between the auxiliary variable $Y$ and the Higgs field $f(r)$. It guarantees that the right-hand side of the Bogomol'nyi equations is real, continuous, and smooth, but it does not by itself constitute a proof for the existence of vortex solutions satisfying the boundary conditions.

\section{Explicit Bogomol'nyi equations and numerical solutions }\label{sec5}

Having established the generic conditions on the constitutive parameter space in Sec.~\ref{sec4}, we now construct explicit BPS equations and field profiles for several representative choices of $\sigma$. Specifically, the governing Bogomol'nyi equations are given by Eqs.~\eqref{eq:f'} and~\eqref{eq:findX} (or, equivalently, Eq.~\eqref{eq:constitutive}), with the corresponding potential determined via Eq.~\eqref{eq:V}. For selected values of $\sigma$, we obtain the auxiliary variable $Y(f)$---along with, consequently, the potential $V(f)$ and the gauge field derivative $a'/r$---in closed form. Since the resulting coupled first-order system admits no closed-form solutions regardless of $\sigma$, the Bogomol'nyi profiles $f(r)$ and $a(r)$ are subsequently computed numerically. This algebraic obstruction is not unique to the Kruglov model: even within standard Born-Infeld-type gauge theories, exact Bogomol'nyi reductions are
known to be difficult to obtain in general, occurring only for distinguished sectors compatible with self-duality~\cite{Yang:2000uj,Xu:2026bdw}.

\subsection{Maxwell and Born-Infeld limits}\label{sec5:maxwellBI}

To validate the general setup, we first examine two paradigmatic benchmark cases corresponding to $\sigma=1$ and $1/2$. Setting $\sigma=1$ recovers the standard canonical Maxwell-Higgs framework, for which the constitutive map and the corresponding Bogomol'nyi equations simplify to:
\begin{subequations}
\begin{align}
    Y &= 1+\frac{\beta e^2}{2}\left(f^2-1\right)^2,\\
     \frac{a'}{r} &= \,e^2\left(f^2-1\right),\\
    V(f) &= \frac{e^2}{2}\left(f^2-1\right)^2.
\end{align}
\end{subequations}
We obtain the usual Mexican-hat potential.

\begin{figure}[t]
    \centering
    \includegraphics[width=0.6\linewidth]{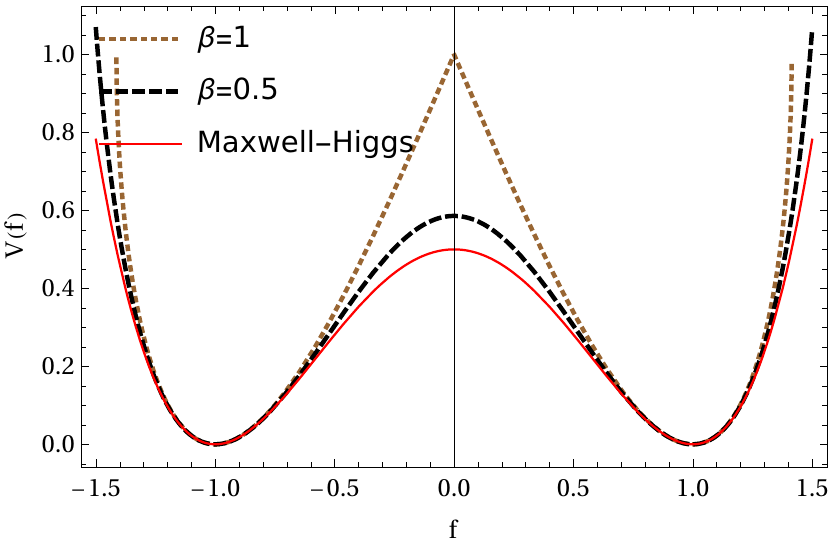}
    \caption{$V(f)$ for the Born--Infeld case
    ($\sigma=1/2$) and the Maxwell--Higgs case ($\sigma=1$),
    with $e=1$ and $s_f=+1$. For the BI case, increasing $\beta$ raises the potential at $f=0$. The potential remains regular for $\beta<1$, while $\beta=1$ represents the limiting case in which the Born--Infeld potential reaches its maximum admissible value at $f=0$.}
    \label{fig:Vsigma1p2}
\end{figure}

Conversely, choosing $\sigma=1/2$ corresponds to the non-linear Born-Infeld gauge dynamics, yielding:
\begin{subequations}
\begin{align}
    Y &= \frac{1}{1-\beta e^2\left(f^2-1\right)^2},\\
    \frac{a'}{r} &= \,e^2\,\frac{\left(f^2-1\right)}{\sqrt{1-\beta e^2\left(f^2-1\right)^2}},\label{aBI}\\
    V(f) &= \frac{1}{\beta}\left[1-\sqrt{1-\beta e^2\left(f^2-1\right)^2}\right],\label{VBI}
\end{align}
\end{subequations}
Note that Eqs.~\eqref{aBI}-\eqref{VBI} exactly reproduce the Bogomol'nyi equation for the gauge field and the corresponding potential for Abelian Born--Infeld--Higgs strings in Ref.~\cite{Yang:2000uj}. These are also consistent with the magnetically-sourced Born-Infeld vortex construction of~\cite{LinYang2003}. The behavior of the potential is shown in Fig.~\ref{fig:Vsigma1p2}.


\subsection{The $\sigma=3/4$ case}
\label{sec5:threefourths}

Beyond the standard benchmarks, the choice $\sigma=3/4$ serves as a representative non-canonical model wherein the constitutive map can still be inverted analytically. Solving Eq.~\eqref{eq:constitutive} directly for the physically admitted branch of $Y$ yields:
\begin{align}
    Y &= 1 + \frac{2}{9}\beta e^2\left(f^2-1\right)^2 \nonumber \\
    &\quad \times
    \left[
    \sqrt{\beta^2 e^4\left(f^2-1\right)^4+9}
    +\beta e^2\left(f^2-1\right)^2
    \right].
    \label{eq:Y34}
\end{align}

It is convenient to introduce the auxiliary function
\begin{equation}
    U(f) =
    \sqrt{\beta^2 e^4\left(f^2-1\right)^4+9}
    +\beta e^2\left(f^2-1\right)^2.
\end{equation}

The gauge-field Bogomol'nyi equation and the corresponding potential can then be written in closed form as
\begin{subequations}
\begin{align}
    \frac{a'}{r}
    &= e^2\left(f^2-1\right)
    \left(
    \frac{2}{9}\beta e^2\left(f^2-1\right)^2 U+1
    \right)^{1/4},\\
    V(f)&= \frac{1}{\beta} \left[\frac{1}{2}\left(\frac{2}{9}\beta e^2\left(f^2-1\right)^2 U+1\right)^{3/4} \right. \nonumber\\
 &\qquad\left.-\frac{3}{2\left(\frac{2}{9}\beta e^2\left(f^2-1\right)^2 U+1\right)^{1/4}}+1\right].
\end{align}
\end{subequations}

To examine the weakly nonlinear regime, we expand these expressions for small $\beta$. This gives
\begin{subequations}
\begin{align}
    \frac{a'}{r}
    &= e^2\left(f^2-1\right)
    +\frac{1}{6}\beta e^4\left(f^2-1\right)^3 \nonumber\\
    &\quad
    +\frac{1}{72}\beta^2e^6\left(f^2-1\right)^5
    +\mathcal{O}\left(\beta^3\right),\\
     V(f)
    &= \frac{1}{2}e^2\left(f^2-1\right)^2
    +\frac{1}{24}\beta e^4\left(f^2-1\right)^4 \nonumber\\
    &\quad
    +\frac{1}{432}\beta^2e^6\left(f^2-1\right)^6
    +\mathcal{O}\left(\beta^3\right).
\end{align}
\end{subequations}
Thus, in the limit $\beta\to0$, both gauge field Bogomo'nyi equation and the potential smoothly recover their Maxwell--Higgs counterparts.

\begin{figure}[t]
    \centering
    \includegraphics[width=0.6\linewidth]{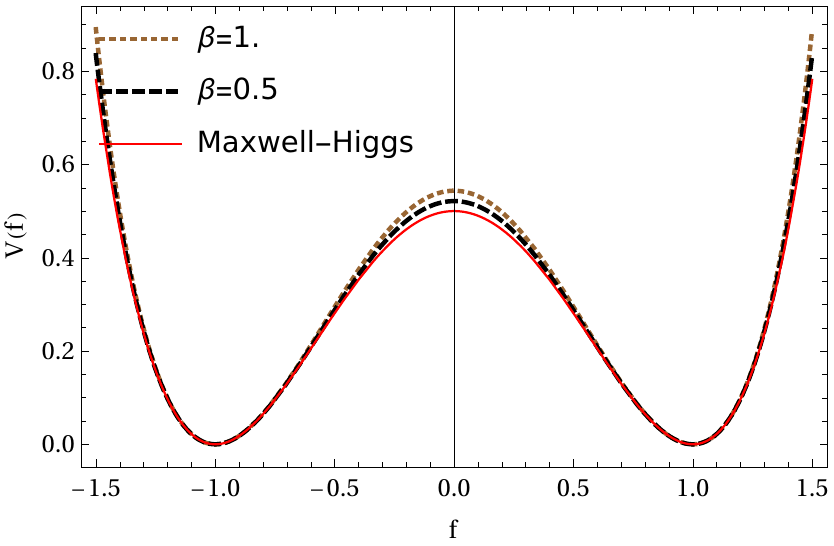}
    \caption{$V(f)$ for the $\sigma=3/4$ case with $e=1$. Increasing $\beta$ enhances the potential at $f=0$.
    In contrast to the BI case, the potential remains regular for $\beta=1$.}
    \label{fig:Vsigma3p4}
\end{figure}

Fig.~\ref{fig:Vsigma3p4} shows the potential for representative values $\beta\in\{0.5,1.0\}$ with $e=1$. Increasing $\beta$ raises the value of the potential at the vortex core, while the potential
remains finite and regular for these parameter values.

\subsection{The $\sigma=5/4$ case}\label{sec5:fivefourths}

For values of $\sigma > 1$, the constitutive mapping exhibits higher-algebraic complexity. In particular, setting $\sigma=5/4$ transforms Eq.~\eqref{eq:constitutive} into a depressed cubic equation for $u \equiv \sqrt{Y}$:
\begin{equation}
u^3-u-C(f)=0.
\end{equation}
Its unique real branch consistent with $u\to1$ as $C\to0$ is
\begin{equation}
u(C) =
\begin{cases}
\dfrac{2}{\sqrt3}\cos\!\left[\dfrac13\arccos\!\left(\dfrac{3\sqrt3}{2}C\right)\right],\\
  0\le C\le C_c,\\ \\[10pt]
\sqrt[3]{\dfrac{C}{2}+\sqrt{\dfrac{C^2}{4}-\dfrac1{27}}}
+\sqrt[3]{\dfrac{C}{2}-\sqrt{\dfrac{C^2}{4}-\dfrac1{27}}},\\  C>C_c,
\end{cases}
\label{eq:u54}
\end{equation}
with $C_c\equiv2\sqrt3/9$, and $Y=u(C(f))^2$. Although Eq.~\eqref{eq:u54}
is expressed in two algebraically distinct forms -- trigonometric and Cardano radical, depending on the sign of the cubic discriminant -- the
physical branch itself is smooth and single-valued throughout $0\le f\le1$, as guaranteed by the monotonicity of $\Phi\left(Y;\sigma\right)$.

The gauge and the potential are given by
\begin{subequations}
\begin{align}
\frac{a'}{r} &= e^2\left(f^2-1\right)Y^{-1/4},\\
V(f) &= \frac{1}{\beta}\left[\frac32 Y^{5/4}-\frac52 Y^{1/4}+1\right],
\end{align}
\end{subequations}
which expand, for small $\beta$, as
\begin{subequations}
\begin{align}
\frac{a'}{r} &= e^2\left(f^2-1\right)-\frac{1}{10}\beta e^4\left(f^2-1\right)^3+O\left(\beta^2\right),\\
V(f) &= \frac12 e^2\left(f^2-1\right)^2-\frac{1}{40}\beta e^4\left(f^2-1\right)^4+O\left(\beta^2\right),
\end{align}
\end{subequations}
correctly recovering the Maxwell-Higgs potential as $\beta\to0$.

\begin{figure}[t]
    \centering
    \includegraphics[width=0.6\linewidth]{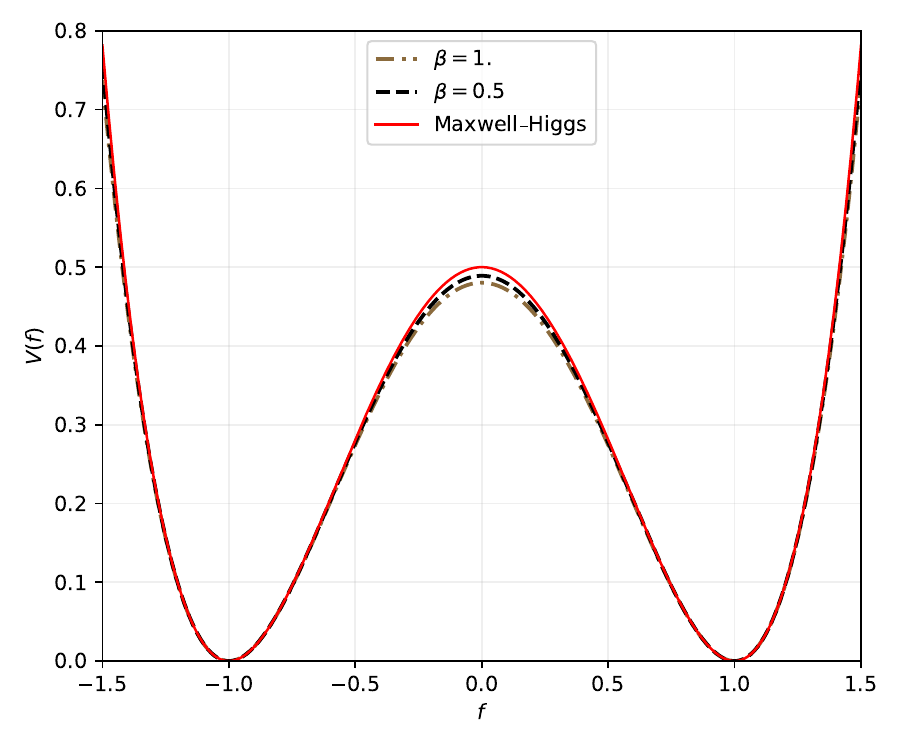}
    \caption{$V(f)$ for the $\sigma=5/4$ case with $e=1$. For the representative value $\beta=0.1$, the potential
    retains the characteristic Mexican-hat shape of the Maxwell--Higgs theory, with its global minimum at $f=1$.}
    \label{fig:Vsigma5p4}
\end{figure}

\subsection{The $\sigma=3/2$ case}
\label{sec5:threehalves}

Another analytically tractable model in the regime $\sigma > 1$ corresponds to $\sigma=3/2$. Here, the constitutive relation reduces remarkably to a quadratic polynomial in $Y$. Selecting the positive, non-singular branch ($Y > 0$) leads to the closed-form expression:
\begin{equation}
    Y =
    \frac{1}{6}
    \left[
    \sqrt{3}
    \sqrt{4\beta e^2\left(f^2-1\right)^2+3}
    +3
    \right].
    \label{eq:Y32}
\end{equation}

The gauge equation and the potential then follow directly from Eqs.~\eqref{eq:V} and~\eqref{eq:a'}, respectively:
\begin{subequations}
\begin{align}
\frac{a'}{r}&= e^2\left(f^2-1\right)Y^{-1/2},\\
    V(f)
    &= \frac{1}{\beta}
    \left[
    2Y^{3/2}-3Y^{1/2}+1
    \right],
\end{align}
\end{subequations}
Upon substituting Eq.~\eqref{eq:Y32}, the former can be written explicitly as
\begin{equation}
    \frac{a'}{r}
    =
    \frac{\sqrt{6}\,e^2\left(f^2-1\right)}
    {\sqrt{
    \sqrt{12\beta e^2\left(f^2-1\right)^2+9}+3
    }}.
\end{equation}

\begin{figure}[t]
    \centering
    \includegraphics[width=0.6\linewidth]{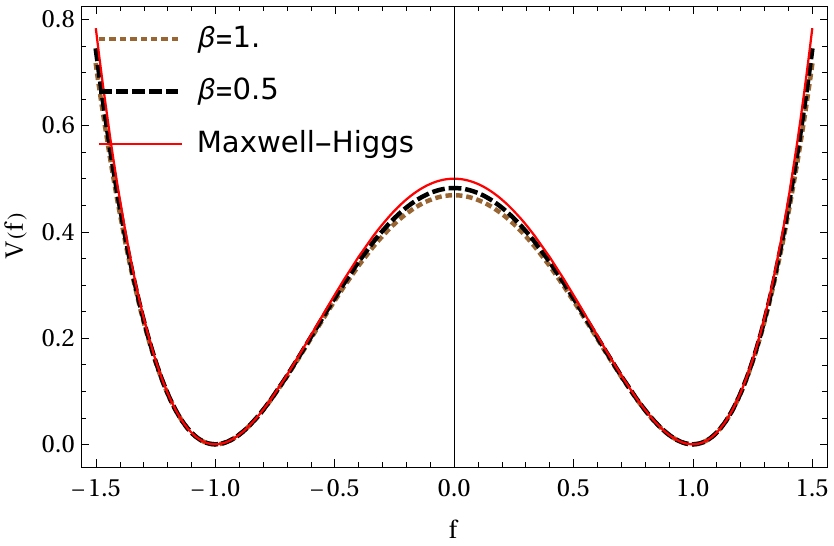}
    \caption{$V(f)$ for the $\sigma=3/2$ case with $e=1$. For the representative value $\beta=0.1$, the potential
    retains the characteristic Mexican-hat shape of the Maxwell--Higgs theory, with its global minimum at $f=1$.}
    \label{fig:Vsigma3p2}
\end{figure}

Expanding for small $\beta$ gives
\begin{subequations}
\begin{align}
 \frac{a'}{r}
    &= e^2\left(f^2-1\right)
    -\frac{1}{6}\beta e^4\left(f^2-1\right)^3 \nonumber\\
    &\quad
    +\frac{7}{72}\beta^2e^6\left(f^2-1\right)^5
    +\mathcal{O}\left(\beta^3\right),\\
    V(f)
    &= \frac{1}{2}e^2\left(f^2-1\right)^2
    -\frac{1}{24}\beta e^4\left(f^2-1\right)^4 \nonumber\\
    &\quad
    +\frac{7}{432}\beta^2e^6\left(f^2-1\right)^6
    +\mathcal{O}\left(\beta^3\right).
\end{align}
\end{subequations}
Again, the $\beta\to0$ limit recovers the standard Maxwell--Higgs potential and the corresponding gauge-field Bogomol'nyi equation. The resulting potential is shown in Fig.~\ref{fig:Vsigma3p2}.

\subsection{The $\sigma=2$ case}\label{sec5:two}

Finally, setting $\sigma=2$ pushes the constitutive dynamics further into the non-linear regime $\sigma > 1$, transforming Eq.~\eqref{eq:constitutive} into the complete cubic equation
\begin{equation}
    Y^3 - Y^2 - C(f) = 0, \qquad C(f) \equiv \frac{\beta e^2}{4}\left(f^2-1\right)^2.
\end{equation}
Unlike the $\sigma=5/4$ scenario, the discriminant of this cubic polynomial, $\Delta = -C\left(27C+4\right)$, remains strictly negative for all physical configurations $C \ge 0$. Consequently, the equation admits a unique real root across the entire domain, removing any need for trigonometric branch switching. 

Introducing the auxiliary function
\begin{equation}
    W(f) \equiv \sqrt{C(f)\left(27C(f) + 4\right)},
\end{equation}
the physical root for $Y$ is expressed in closed Cardano form as
\begin{equation}
    Y = \frac{1}{3} + \sqrt[3]{\frac{C}{2} + \frac{1}{27} + \frac{\sqrt{3}}{18}W} + \sqrt[3]{\frac{C}{2} + \frac{1}{27} - \frac{\sqrt{3}}{18}W}.
    \label{eq:Y2}
\end{equation}
As guaranteed by the monotonicity of the mapping $\Phi(Y;\sigma)$, Eq.~\eqref{eq:Y2} remains strictly real, positive, and smooth for all $0 \le f \le 1$, without generating complex artifacts.

The gauge equation and the potential then take the forms
\begin{equation}
\frac{a'}{r} = \frac{e^2\left(f^2-1\right)}{Y},\qquad V(f) = \frac{1}{\beta}\left(3Y^2 - 4Y + 1\right).
\end{equation}
In the small-$\beta$ expansion, these expressions yield
\begin{align}
\frac{a'}{r} &= e^2\left(f^2-1\right) - \frac{1}{4}\beta e^4\left(f^2-1\right)^3 + \mathcal{O}\left(\beta^2\right),\\
    V(f) &= \frac{1}{2} e^2\left(f^2-1\right)^2 - \frac{1}{16}\beta e^4\left(f^2-1\right)^4 + \mathcal{O}\left(\beta^2\right),
\end{align}
which smoothly reduce to the standard Maxwell-Higgs theory in the limit $\beta \to 0$.

\subsection{Numerical solutions}

For the representative values of $\sigma$ considered above, the constitutive relation can be solved explicitly to obtain $Y(f)$, from which the corresponding potential $V(f)$ and gauge-field
Bogomol'nyi equation $a'/r$ follow. The remaining first-order equation for the Higgs profile, Eq.~\eqref{eq:f'}, together with the gauge-field equation, do not admit a closed-form
solution for the vortex profiles. We therefore solve the resulting first-order system numerically.

In this analysis, we restrict ourselves to
parameter values for which the constitutive relation and the resulting potential remain real throughout the relevant field range $0\leq f\leq1$. The cases $\sigma=1/2$, $3/4$, $1$, and $3/2$ all satisfy this requirement for the parameter choice considered below.

\begin{figure}
    \centering
    \includegraphics[width=0.6
    \linewidth]{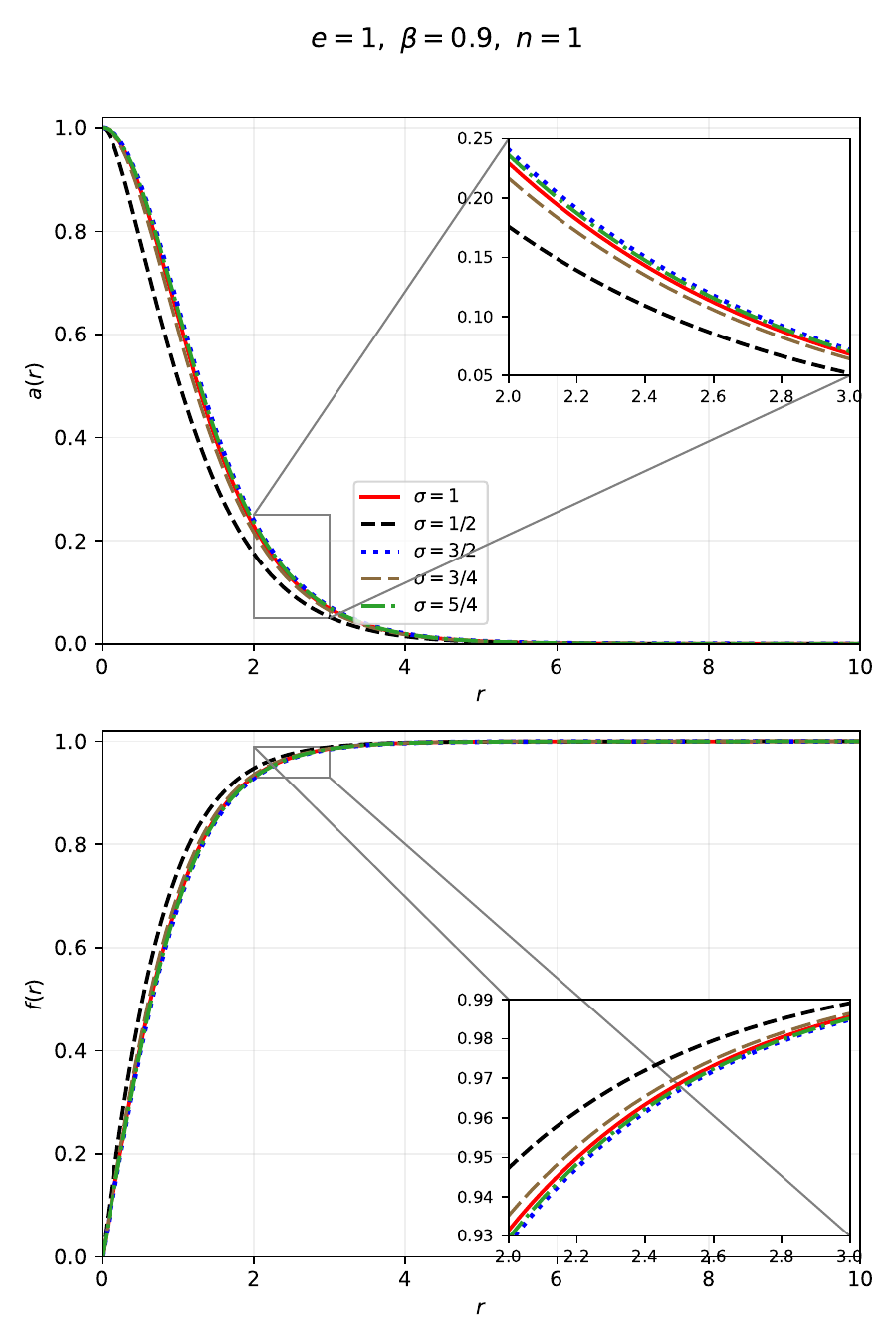}
    \caption{Numerical BPS vortex profiles for
    $\sigma=1/2$, $3/4$, $1$, $3/4$ and $3/2$, with $e=1$ and $\beta=0.9$. The Higgs profile $f(r)$ increases from the vortex core towards its vacuum value, while the gauge profile $a(r)$ decreases from $n$ to zero. The different values of $\sigma$ produce distinct characteristic vortex-core widths.}
    \label{fig:numSolbeta0.9}
\end{figure}

Fig.~\ref{fig:numSolbeta0.9} presents the numerical BPS profiles for these four cases with $e=1$ and $\beta=0.9$. The profiles satisfy the
boundary conditions, Eq.~\eqref{boundary}, and exhibit the expected monotonic behavior: $f(r)$ increases from the vortex core towards its vacuum value, while $a(r)$ decreases from $n$ towards zero. The nonlinear parameter $\beta$ controls the quantitative differences among the profiles, which become more pronounced as the nonlinearity increases.

For $\beta=0.9$, the gauge-field profile $a(r)$ decreases most rapidly for $\sigma=1/2$, followed by $\sigma=3/4$, $\sigma=1$, and $\sigma=3/2$. Correspondingly, the Higgs profile $f(r)$ increases in the same order, with the $\sigma=1/2$ solution approaching its vacuum value most rapidly and the $\sigma=3/2$ solution most slowly. Thus, the different constitutive structures associated with $\sigma$ lead to distinct characteristic widths of the vortex core, despite all solutions sharing the same BPS tension $\mu_{\mathrm{BPS}}=2\pi n$.



\section{Conclusions}\label{sec6}

The construction of Bogomol'nyi equations through first-order formalisms has been extensively developed for generalized field theories beyond the standard Abelian--Higgs model. These approaches provide a systematic way of reducing the second-order field equations to a set of first-order equations while preserving the topological character of the Bogomol'nyi bound. Their application to nonlinear electrodynamics, however, becomes considerably more involved. In particular, for polynomial nonlinear electrodynamics, the nonlinear structure of the gauge-field sector leads to a nontrivial stress tensor and prevents the straightforward use of the standard Bogomol'nyi completion of the static energy. This raises the question of whether a systematic first-order construction can be formulated for polynomial NLED--Higgs systems while retaining a genuine Bogomol'nyi bound.

In this work, we show that such a construction is indeed possible for the Kruglov nonlinear electrodynamics coupled to the Abelian--Higgs
model. Kruglov electrodynamics serves as a natural polynomial representative of NLEDs---interpolating between Maxwell ($\sigma=1$), Born-Infeld ($\sigma=1/2$), and exponential electrodynamics ($\sigma \to \infty$). By employing the stressless conditions, we derive a system of first-order Bogomol'nyi equations without specifying the Higgs potential \textit{a priori}. The resulting equations determine the potential and the gauge-field constitutive relation simultaneously, and the
corresponding BPS energy density reduces to a total derivative. As a consequence, the string tension is purely topological, $\mu_{\rm BPS}=2\pi n$, independent of both $\sigma$ and $\beta$. 

For generic values of $\sigma$, the first-order construction leads to an implicit algebraic relation between the auxiliary variable $Y$ and the Higgs profile $f$. Rather than attempting to solve this relation separately for each value of $\sigma$, we introduced a constitutive map $\Phi\left(Y;\sigma\right)$ and analyzed its domain, monotonicity, and range. This analysis reveals a qualitative distinction at $\sigma=1/2$. For $\sigma>1/2$, the constitutive map is strictly monotonic and unbounded on the physical domain $Y>1$, whereas for $0<\sigma<1/2$ it possesses a finite maximum. The marginal case $\sigma=1/2$ is bounded, with $\sup\Phi=1$. These properties lead to explicit conditions on the coupling $\beta$ for the existence of an admissible constitutive branch connecting the vortex core to the vacuum. We emphasize, as already noted in Sec.~\ref{sec4}, that this criterion guarantees only the smoothness of the algebraic constitutive relation $Y(f)$.

A broader implication of our analysis is that the existence problem for the first-order vortex equations can be separated from the subsequent
boundary-value problem for the vortex profiles. Once the stressless conditions reduce the gauge and Higgs sectors to an algebraic constitutive
relation, the admissibility of the Bogomol'nyi branch can be assessed directly from the domain, monotonicity, and range of the constitutive map.
This provides a simple diagnostic for determining whether a proposed nonlinear gauge sector can support a regular BPS branch before solving the
field equations numerically. In this sense, the constitutive-map approach offers a general strategy for extending first-order vortex constructions to broader classes of nonlinear electrodynamics.

Guided by this criterion, we obtained the constitutive relation $Y(f)$, and consequently the potential $V(f)$ and gauge equation $a'/r$, for six representative values of $\sigma$, spanning a hierarchy of increasing algebraic complexity: linear ($\sigma=1$), quadratic in $Y$ ($\sigma=3/2$), quadratic in $\sqrt Y$ ($\sigma\in\{1/2,3/4\}$), and cubic in $\sqrt Y$ or $Y$ itself ($\sigma\in\{5/4,2\}$), the latter two requiring, respectively, a trigonometric/Cardano branch structure and a single-valued Cardano root.
In every case the potential and gauge equation were shown to reduce smoothly to their Maxwell--Higgs counterparts as $\beta\to0$, and the
corresponding vortex profiles $f(r)$, $a(r)$ were obtained numerically.

We should also comment on the physical consistency of the parameter space considered in this work. Throughout the analysis, we have fixed
$\gamma=\beta/\left(2\sigma\right)$. Recent analysis by Russo and Townsend~\cite{Russo:2024kto,Russo:2024xnh} have established stringent constraints on nonlinear electrodynamics arising from causality, characteristic propagation, and energy conditions. 
More explicitly, $1/2\leq\sigma<1$ and $\gamma=\beta/\left(2\sigma\right)$ are required to satisfy both the weak and strong field causality. While these causality conditions satisfy the weak and strong energy conditions, the converse is not necessarily true.
These constraints are derived by requiring the underlying NLED theory to remain consistent for generic electromagnetic backgrounds. By contrast, the vortex configurations studied here probe only the purely magnetic sector, for which $\mathcal{G}=0$, and therefore do not provide a test of the full characteristic structure of the
theory. Consequently, the existence of regular BPS vortex solutions within the constitutive construction should not be regarded as a proof
of the causal or energetic consistency of the complete Kruglov--Higgs theory. In particular, the solutions with $\sigma>1$ considered in this work should be viewed as formal extensions of the BPS constitutive construction until the corresponding characteristic and consistency conditions are established for the full theory. A complete analysis of these issues, including the effects of the Higgs sector on the propagation of electromagnetic perturbations, is beyond the scope of the present work. 

These results suggest several directions for further investigation. The BPS Lagrangian~\cite{Atmaja:2015umo,Atmaja:2018cod} and on-shell methods~\cite{Atmaja:2014fha} have already been employed to construct first-order equations for $SU(2)$ Yang--Mills--Higgs monopoles and dyons using auxiliary algebraic relations of essentially the same type as Eq.~\eqref{eq:findX}. It would therefore be natural to replace the Yang-Mills or non-Abelian Born--Infeld gauge sector in these constructions by the full non-Abelian Kruglov $\sigma$-family and examine whether the existence criterion, Eq.~\eqref{eq:existence}, leads to analogous restrictions on the admissible parameter space. Another promising direction is the construction of Bogomol'nyi equations for gauge fields governed by exponential electrodynamics~\cite{Hendi:2012zz,Hendi:2013dwa}, which arises as the $\sigma\to\infty$ limit of the Kruglov Lagrangian. Finally, in light of the string-theoretic origin of Born--Infeld-type actions as effective D-brane dynamics~\cite{Fradkin:1985qd,Leigh:1989jq,Seiberg:1999vs}, it would be interesting to investigate whether the Kruglov $\sigma$-family, and the vortex solutions constructed here in particular, admit a corresponding brane-theoretic interpretation, potentially connecting them with the Dirac--Born--Infeld cosmic strings studied in~\cite{Ramadhan:2025rxj}.

More concretely, the Kruglov Lagrangian's power-law kinetic structure, in contrast to the quadratic Maxwell or single-square-root Born--Infeld kinetic terms, renders the resulting constitutive equations genuinely polynomial in $Y$ for rational $\sigma$. Certifying polynomial non-negativity and establishing positivity conditions for high-degree polynomial expressions are central problems in sum-of-squares (SOS) and polynomial optimization, with substantial applications in engineering, including control theory through Lyapunov-function synthesis for nonlinear stability certification, robotics, and power-systems engineering~\cite{Parrilo2003,Lasserre2001}. The polynomial reductions obtained in Sec.~\ref{sec5} may therefore be viewed as low-dimensional physical instances of this broader class of algebraic feasibility problems. More generally, our constitutive-map analysis suggests that, in suitable nonlinear systems, the admissibility and branch structure of an underlying algebraic relation can be assessed before solving the associated differential boundary-value problem. This viewpoint may provide a useful connection between first-order constructions in nonlinear field theory and broader problems in nonlinear constitutive modeling, polynomial optimization, and constrained dynamical systems.

\section*{Acknowledgement}
We thank Naufal Athaullah for fruitful discussions. HSR is supported by Hibah PUTI Q1 UI No.~PKS196/UN2.RST /HKP.05.00/2025 and expresses sincere gratitude to the Ministry of Higher Education, Science, and Technology of the Republic of Indonesia for their support through Hibah Fundamental DIKTI No.~PKS-178/UN2.RST/HKP.05.00/2026. AI assistance was utilized to support parts of the numerical calculations and provided language editing. The authors retain full responsibility for the final mathematical results and text.

\begin{appendix}

\end{appendix}


\end{document}